\documentclass[pdflatex,sn-mathphys-num]{sn-jnl}

\usepackage{graphicx}%
\usepackage{multirow}%
\usepackage{amsmath,amssymb,amsfonts}%
\usepackage{amsthm}%
\usepackage{mathrsfs}%
\usepackage[title]{appendix}%
\usepackage{xcolor}%
\usepackage{textcomp}%
\usepackage{manyfoot}%
\usepackage{booktabs}%
\usepackage{algorithm}%
\usepackage{algorithmicx}%
\usepackage{algpseudocode}%
\usepackage{listings}%
\usepackage{amsmath,amssymb,amsthm}
\usepackage{graphicx}
\usepackage{tikz}
\usetikzlibrary{arrows.meta, positioning}
\usepackage{hyperref}
\usepackage{enumitem}
\usepackage{booktabs}
\usepackage{multirow}
\usepackage{fvextra}
\usepackage{hyperref}
\usepackage{natbib}
\usepackage{tcolorbox}
\tcbuselibrary{listings, breakable}
\usepackage[table]{xcolor}
\usepackage{longtable}

\theoremstyle{thmstyleone}%
\theoremstyle{thmstyletwo}%
\newtheorem{remark}{Remark}%

\theoremstyle{thmstylethree}%

\begin{document}

\title[Structured but Fragile]{Structured but Fragile: On the Limits of LLMs in Cybersecurity Decision-Making}


\author[1]{\fnm{Pasquale} \sur{Malacaria}}\email{p.malacaria@qmul.ac.uk}
\equalcont{These authors contributed equally to this work.}

\author*[2]{\fnm{Yunxiao} \sur{Zhang}}\email{y.zhang12@exeter.ac.uk}
\equalcont{These authors contributed equally to this work.}

\affil[1]{\orgdiv{School of Electronic Engineering and Computer Science}, \orgname{Queen Mary University of London}, \orgaddress{ \city{London}, \country{UK}}}

\affil[2]{\orgdiv{Department of Computer Science}, \orgname{University of Exeter}, \orgaddress{\city{Exeter}, \country{UK}}}


\abstract{Large language models (LLMs) are increasingly used in cybersecurity workflows, yet it remains unclear whether they can perform structured security reasoning or merely rely on superficial cues and prior knowledge.
We study this question in the context of defence selection over attack graphs derived from real-world threat scenarios, including ransomware, supply-chain compromise, cloud abuse, Kubernetes attacks, POS malware, and ICS/OT intrusion. Given a budget constraint, LLMs must select security controls to minimise attacker success. We compare their strategies against each other and against a game-theoretic optimization baseline used as a normative reference for structured reasoning.
Our results show that LLMs exhibit conditional competence. When explicit attack-graph structure is provided, they often produce coherent strategies close to the optimization baseline. 
However, their capabilities are fragile. 
LLM behaviour becomes increasingly fragile with graph complexity and is highly sensitive to framing.
Small prompt changes can substantially alter rankings, and merely relabeling a poor strategy as “optimal” dramatically improves its evaluation. We further observe a non-monotonic relationship between formal risk and LLM judgement: strategies closest to the optimum are not necessarily ranked highest by LLM evaluators.
To further probe reasoning ability, we ask LLMs to generate solvers for the same optimization problem. While the generated implementations recover the correct high-level formulation, they scale poorly compared to a purpose-built solver.
Overall, our findings show that LLMs can approximate structured cybersecurity reasoning under controlled representations, but do not apply it robustly. This has important implications for the design and evaluation of AI-assisted security decision-support systems.}


\maketitle

\section{Introduction}

Large language models (LLMs) are rapidly being adopted in cybersecurity, supporting tasks such as threat analysis, mitigation planning, and incident response. A natural next step is to use them as decision-makers: given a threat scenario and a set of defensive options, can an LLM identify an effective defence strategy?

This question is challenging because security decisions are inherently structured. Defenders must allocate limited resources across multiple controls, reason about attacker progression, and prioritise interventions that disrupt critical attack paths. These problems are naturally captured by attack graphs, where nodes represent attacker states and edges represent attack steps. Effective decision-making requires reasoning over this structure, not just recalling best practices.

A challenge is that an LLM may produce a strong defence because it genuinely reasons about attack paths, or because it recognises familiar controls (e.g., MFA, segmentation) and responds with generic security intuition. Similarly, evaluations may reflect true quality or be influenced by superficial cues such as naming or framing. For security applications, this distinction is critical.

In this paper, we study the capabilities and limitations of frontier LLMs as cybersecurity decision-makers. We consider seven attack-graph scenarios derived from real-world threats, including ransomware, supply-chain compromise, cloud abuse, POS attacks, Kubernetes compromise, and ICS/OT intrusion. For each scenario and budget level, LLMs are asked to select defence portfolios. These portfolios (and other strategies) are then evaluated by the LLMs under controlled conditions. We include a game-theoretical solution as an independent baseline, which serves as a normative reference for structured decision-making under a well-defined objective.

Our results show that LLMs can produce strong strategies and can discriminate between good and poor ones. 
At the same time, we identify significant vulnerabilities. LLM evaluations are sensitive to framing and naming: a poor strategy labeled “optimal” may be ranked by LLMs above genuinely strong strategies. We also observe moderate self-preference effects and outlier cases where models prioritise domain knowledge over explicit threat structure. Again, these are concerning aspects in a security context.

To further explore their reasoning ability, we ask LLMs to generate code that solves this type of defence-selection problems. We evaluate these generated solvers on 36 randomly generated attack graphs of increasing size. The resulting solutions are optimal but struggle to scale to large graphs.

Taken together, these findings reveal a consistent pattern. LLMs can approximate structured reasoning when the attack-graph representation and objective are explicit, producing strategies that align with a principled optimization baseline. However, this capability is not robust: it degrades under perturbations such as framing and abstraction, and can be overridden by general domain knowledge. We therefore characterise LLM behaviour as conditionally competent, rather than reliably optimal, in cybersecurity decision-making tasks.

Our contribution is a controlled evaluation framework for studying structured LLM cybersecurity decision-making, together with an empirical characterisation of when such reasoning succeeds, when it fails, and how these failures arise. 

\subsection{Contributions}

This paper makes the following contributions:

\begin{itemize}
\item  \textbf{A controlled methodology for evaluating structured LLM cybersecurity decision-making.}
We provide a systematic evaluation of frontier LLMs on defence-selection tasks over attack graphs derived from seven realistic cybersecurity scenarios (e.g., ransomware, cloud, supply chain, POS, Kubernetes, ICS/OT).

\item \textbf{Normative benchmark for structured reasoning.}
We use a game-theoretic optimization baseline as a controlled reference that encodes explicit attacker–defender trade-offs, enabling evaluation of whether LLMs follow structured reasoning without assuming ground-truth labels.

\item \textbf{Conditional competence under structure.}
We show that LLMs can produce coherent strategies and approximate high-quality solutions when the attack graph is small and the structure is explicit, often aligning with the optimization baseline. 

\item \textbf{Framing and naming.}
We detect framing and naming effects, showing that evaluation outcomes can be significantly altered by presentation, including cases where weak strategies are ranked higher than optimal ones under misleading labels.

 \item \textbf{Non-monotonic evaluation behaviour.} We identify cases where LLM evaluators fail to distinguish near-optimal strategies, ranking mathematically close solutions substantially below the optimum. This reveals a mismatch between formal risk minimisation and LLM judgement, particularly on complex graphs.

\item \textbf{LLM-generated optimization code.}
We show that LLMs can generate solvers that recover the correct optimization approach, but that these implementations are not sophisticated and do not scale compared to the purpose-built game-theoretic baseline.

\item \textbf{Implications for security workflows.} We derive practical recommendations for using LLMs in security decision-support systems, including anonymisation, avoidance of suggestive framing, and the necessity of explicit structural representations.
\end{itemize}

\section{Related Work}

\subsection{Cybersecurity Decision-Making}

Cybersecurity decision-making concerns analysing risks, allocating defensive resources, and selecting security controls to protect organisational assets while balancing security needs with business operations under uncertainty \cite{ainslie2023cyber,ZHANG2025104153,tifs_2026,fielder2018risk}. A key problem in this area is cybersecurity investment: decision-makers must determine how much to invest, which assets to prioritise, and which controls to deploy in order to maximise protection while limiting cost and disruption \cite{zhang2021bayesian,fielder2016decision,zhang2023keep,khouzani2019scalable,sawik2013selection,anderson2010security,sawik2026cybersecurity}.

Existing approaches can be broadly divided into \textit{qualitative} and \textit{quantitative} methods. Qualitative methods are typically scenario-based, relying on expert judgement to assess threats, risks, and controls using relative scales or simple categories \cite{NIST-SP800-30}. Quantitative methods instead use numerical estimates of factors such as threat likelihood, incident loss, control cost, and risk reduction \cite{gordon2002economics}.

Recent cybersecurity decision-support research has largely adopted quantitative models, particularly Stackelberg security games, which capture the adversarial interaction between defender and attacker \cite{tambe2011security,cavusoglu2008decision,karwowski2023sequential,korzhyk2011stackelberg,fielder2016decision,zhang2021bayesian,khouzani2016efficient,khouzani2019scalable,zhang2023keep}. In these models, the defender commits to a strategy first, anticipating the attacker’s best response. The environment is often represented as an attack graph, which describes the possible paths an attacker may exploit to reach a target \cite{ou2006scalable,ZDEMIRSNMEZ2022102938,bayesian_attack_graph}. Most relevant to our work is \cite{khouzani2019scalable}, which formulates cybersecurity investment as a multi-objective bi-level Stackelberg game over probabilistic attack graphs and provides an efficient solution. However, accurate estimates of attacker capability, control effectiveness, and potential loss are often unavailable, which makes uncertainty a central challenge \cite{zhang2021bayesian,zhang2023keep,ZHANG2025104153}.

An important motivation for this work is that human cybersecurity experts are expensive and scarce \cite{blavzivc2021cybersecurity}, so we investigate whether recent LLMs can support, or partially replace, expert judgement in cybersecurity decision-making by drawing on broad prior knowledge, pattern recognition, and context-aware recommendation capabilities \cite{levi2025cyberpal,israr2024enhancing}.

\subsection{LLMs for cybersecurity}

The integration of LLMs into cybersecurity operations has emerged as a promising way to augment, or partially replace, human experts in security decision-making. Recent studies have examined how LLMs can draw on domain knowledge, recognise threat patterns, generate context-aware recommendations, and support incident response workflows. 
The effective deployment of LLMs in cybersecurity requires model outputs to be grounded in verifiable, domain-specific knowledge in order to ensure reliability. Recent research introduced several strategies to achieve this.  CyberAlly \cite{CyberAlly} shows that integrating LLMs with cybersecurity knowledge graphs can improve the efficiency and effectiveness of blue teams during incident response, which significantly reduces the risk of hallucination. Another approach to integrating cybersecurity knowledge is retrieval-augmented generation (RAG). CyberRAG \cite{CyberRAG} uses an agentic RAG system to deliver real-time classification, explanation, and structured reporting for cyber-attacks. The adopted agentic architecture allows CyberRAG to have dynamic control flow and adaptive reasoning. 
Similarly, \cite{oh2026automated} introduced an LLM pipeline that maps logs to the MITRE ATT\&CK framework to pinpoint policy gaps, though they caution that high-stakes decisions require human oversight.

Much prior work has focused either on small BERT-style models or on fine-tuning rather than building broader cyber knowledge into the model itself.
PRIMUS is an open-source cybersecurity training suite for LLMs \cite{yu2025primus}, which covers pretraining, instruction fine-tuning, and reasoning distillation. This database allows us to train a cybersecurity LLM end-to-end. Rather than building a general cybersecurity foundation model, SEVENLLM \cite{ji2024sevenllm} focuses on cyber threat intelligence (CTI) analysis, particularly the extraction and generation of structured information from security reports. It covers 28 distinct tasks spanning nearly every aspect of incident report analysis.
These studies have demonstrated the capacity of LLMs to make reasonable security decisions and shown that their performance can be further strengthened through external knowledge sources, such as RAG and knowledge graphs, or through domain-specific training.
The latest \href{https://www.anthropic.com/glasswing}{Project Glasswing} by Anthropic built a frontier model (Claude Mythos) designed to help find and fix software vulnerabilities in critical systems. It aims to give defenders an advantage in the AI-driven cybersecurity era.


These studies demonstrate that modern LLMs can acquire or access substantial cybersecurity knowledge through pre-training, domain-specific fine-tuning, retrieval-augmented generation, and knowledge-graph integration. However, possessing cybersecurity knowledge is different from making structured security decisions under explicit resource and attack-path constraints. Existing work has largely focused on tasks such as threat identification, information extraction, vulnerability analysis, incident response, and recommendation generation, where performance is typically assessed through task accuracy or expert-oriented evaluation.

In contrast, comparatively little attention has been paid to whether frontier LLMs can reason consistently over an explicit cybersecurity decision structure, such as an attack graph, when required to select a portfolio of controls subject to budget constraints. It is also unclear whether apparently strong decisions reflect reasoning over the underlying attack topology or instead arise from familiar control names, domain priors, or evaluation framing.

This work addresses this gap by introducing a controlled evaluation methodology in which LLM-generated defence portfolios are compared against a formal game-theoretic optimisation reference and then subjected to controlled perturbations of framing, naming, semantic information, graph structure, and problem complexity. Our objective is therefore not simply to benchmark cybersecurity knowledge, but to examine when LLMs exhibit structured decision-making behaviour, how closely that behaviour aligns with a formal objective, and under what conditions that alignment breaks down.

\section{Cybersecurity Decision Making}
\label{sec:decision-making}

Here, we present the cybersecurity decision-making problem and formulate it mathematically. We model the cybersecurity investment problem as a Stackelberg security game, using an attack graph to represent the organization’s systems and networks.

Formally, let $G = (V,E)$ (where $V,E$ are the vertices and edges) denote an attack graph, $C$ a set of available controls, and $B$ a budget constraint. The defender's strategy is a subset $s \subseteq C$ satisfying the budget constraint, and $\mathcal{S}(B)$ denotes the set of feasible strategies.
Let $\theta$ denote unknown parameters governing control effectiveness. Given $(G,C,\theta)$, the defender seeks to minimise a risk functional
\[
R(s;\theta),
\]
yielding an optimal strategy:
\[
s^* \in \arg\min_{s \in \mathcal{S}(B)} R(s;\theta).
\]
As $\theta$ is not empirically observable, direct validation of $s^*$ against real-world outcomes is infeasible.

The risk functional $  R(s; \theta)  $ quantifies the defender’s security exposure and is defined as the maximum (worst-case) probability that the attacker reaches the target node when the defender has committed to strategy $  s  $ and the attacker subsequently chooses the best path:
\begin{align}
R(s; \theta) = \max_{\pi \in \Pi(G)} \prod_{e \in \pi} p_e(s; \theta),
\label{eq:R}
\end{align}
where $  \Pi(G)  $ is the set of all paths from the source node (Node 0) to the designated target node in the attack graph $  G = (V, E)  $;
$  p_e(s; \theta) \in [0,1]  $ is the residual success probability of the attacker traversing edge $  e  $ under defence $  s  $.
If no control applicable to edge $  e  $ is selected in $  s  $, then $  p_e(s; \theta) = \pi_e \in [0,1] $ (in this paper, we assume $\pi_e =1$). Otherwise, $  p_e(s; \theta) $ equals to $\pi_e$ multiplied with the effectiveness values of all controls in $s$ applicable to edge $e$. Controls act multiplicatively and independently across edges.
The defender’s problem is therefore the classic min-max Stackelberg formulation:
\begin{align}
    s^* \in \arg\min_{s \in S(B)} R(s; \theta), \label{eq:min-max}
\end{align}
i.e., choose the feasible defence portfolio that minimises the attacker’s best-response success probability.

Please note that throughout this paper,  the term ``\textit{optimal}'' refers to optimality under the specified attack-graph model and its assumed parameters. We do not interpret this solution as an empirically validated real-world ground truth. Rather, the optimisation model provides a controlled normative reference against which LLM decision-making can be compared.

\begin{figure}[h!]
    \centering
    \includegraphics[width=0.35\linewidth]{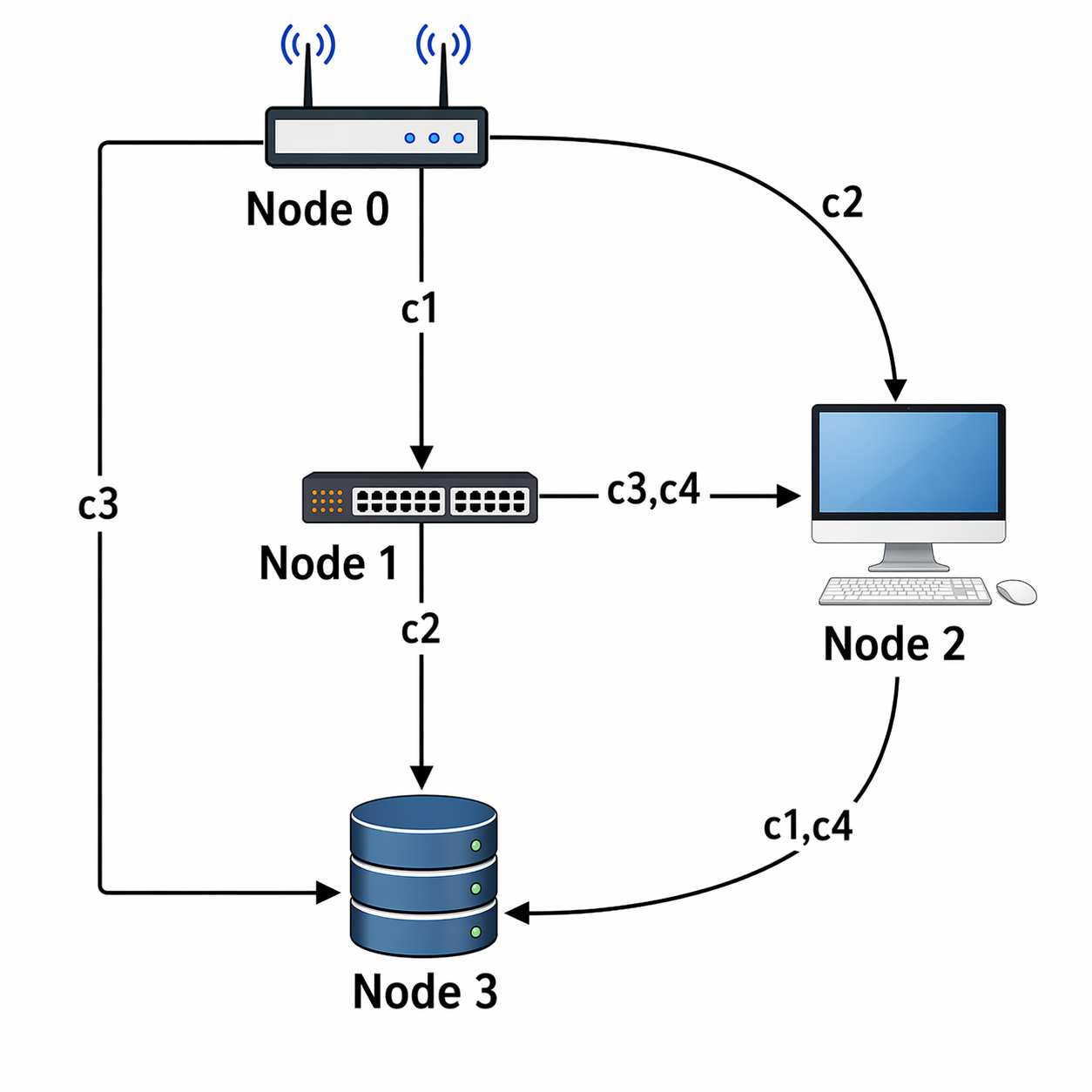}
    \caption{A simple attack graph.}
    \label{fig:attack graph}
\end{figure}

\paragraph{Example} To fix the ideas let's consider the simple attack graph in Fig. \ref{fig:attack graph}. In this graph, node 0 denotes the attack source, e.g., an external network. Nodes 1 and 2 denote intermediate privileged states, for example, a router and a workstation, respectively. Node 3 denotes the attacker’s target, e.g., a database.
An attacker may seek to compromise the database directly through the edge 0 → 3. Alternatively, the attacker may first gain access to either the router via 0 → 1 or the workstation via 0 → 2, and then use that foothold to progress toward the database.

The database is vulnerable to both direct attackers' exploitation (edge 0 → 3) and multi-stage attacks: 0 → 1 → 3, 0 → 2 → 3, 0 → 1 → 2 → 3. Defender can mitigate risk by deploying security controls. There is a set of controls \{c1, c2, c3, c4\}, each representing a technical or administrative safeguard, such as firewalls, user training, etc. The distribution of these controls across the network graph is detailed in Fig. \ref{fig:attack graph}.

Assume that controls c1 to c4 have effectiveness values of 0.4, 0.5, 0.1, and 0.1, respectively. The effectiveness of a security control is defined as the reduction in the attack success probability, $p_e$. For instance, applying control c3 decreases the success probability of edge 0 → 3 by $0.1= 1 \times 0.1$. As mentioned, $\theta$ is not empirically observed, i.e., it is infeasible to determine the exact effectiveness of the controls. However, many quantitative approaches still require numerical inputs in order for the algorithms to compute a solution. There are approaches to address such uncertainties, discussed in \cite{ZHANG2025104153,fielder2018risk}.
The effectiveness values used here are purely illustrative and included to help the reader’s understanding of the decision-making problem.  

Assume a budget of two, meaning that no more than two controls may be selected. Using exhaustive search, we are easy to see that the optimal security portfolio is then $s^* = [c_1, c_3]$, which minimises $R$. However, for large attack graphs, an exhaustive search becomes computationally intractable, since the underlying planning problem is NP-hard \cite{sinha2017review}.

\paragraph{Efficient Game Solution} 
Extensive research has been conducted on efficient methods to find the optimal security portfolio that minimises security risk with large attack graphs \cite{khouzani2019scalable,fielder2016decision,khouzani2016efficient,sawik2013selection}.
Among these, \cite{khouzani2019scalable} introduces a scalable min–max formulation over probabilistic attack graphs to compute the optimal security portfolio. 
\begin{remark}
    The key idea in \cite{khouzani2019scalable} is a mathematical reformulation that reduces the original non-linear bi-level min–max problem in \eqref{eq:R} and \eqref{eq:min-max} to an efficient mixed-integer linear programme (MILP) that standard solvers can handle at scale: the probability of an attack path is originally a product of edge probabilities, but the paper takes logarithms, which converts this into a sum, thereby making the problem linear. It then represents the attacker’s path using edge-selection variables with flow constraints, shows that this attacker problem can be solved exactly as a linear programme, and finally dualises that inner attacker LP and combines it with the defender’s decision variables.
\end{remark}
 
For example, optimal solutions for attack graphs with 20,000 nodes can typically be obtained in less than four minutes \cite{khouzani2019scalable}. Using this solver, all game-theoretical solutions presented in this paper are computed within a few seconds, whereas LLM-generated solutions typically take several minutes per instance.

\section{Methodology}
\label{sec:method}

\subsection{Threat scenarios}
\label{sec:threat_scenarios}
In this paper, we consider the following threat scenarios:
\begin{enumerate}
    \item \textbf{ICS / OT intrusion (IT $\rightarrow$ OT with alternative paths) (6 nodes, 6 edges, 2 paths, 5 controls).} This is a classic branching IT/OT convergence graph used in critical-infrastructure modeling.

    \item \textbf{Double-extortion ransomware campaign (7 nodes, 7 edges, 2 paths, 37 controls).} The attacker gains initial access via remote services, phishing, or exploitation; establishes persistence; steals credentials or escalates privileges; moves laterally; stages data for exfiltration; and finally executes impact through encryption and extortion, or extortion through exfiltration.
    \emph{Source:} \href{https://www.verizon.com/business/resources/T16f/reports/2025-dbir-data-breach-investigations-report.pdf}{Verizon DBIR 2025}.

    \item \textbf{ICS/OT industrial control system attack (15 nodes, 18 edges, 12 paths, 16 controls).} This scenario models an IT-to-OT pivot targeting industrial control systems, from initial access to production or safety impact. It extends the first scenario.

    \item \textbf{Software supply-chain compromise and customer breach (14 nodes, 13 edges, 1 path, 13 controls).} An attacker compromises a software vendor, poisons the build pipeline, and ships a signed malicious update that is installed by customers. The backdoor then enables credential theft, lateral movement, and operational impact. This models SolarWinds-, 3CX-, and XZ-style incidents.
    \emph{Source:} \href{https://attack.mitre.org/campaigns/C0024/}{MITRE ATT\&CK Campaign C0024}.

    \item \textbf{Cloud infrastructure abuse, cryptomining, and data theft (15 nodes, 17 edges, 6 paths, 11 controls).} The attacker exploits cloud IAM misconfigurations, such as over-permissive roles or leaked keys, provisions resources, exfiltrates data, and causes both financial and confidentiality harm.
    \emph{Source:} \href{https://unit42.paloaltonetworks.com/compromised-cloud-compute-credentials/}{Unit 42}.

    \item \textbf{Retail POS compromise, payment-card theft, and cash-out (20 nodes, 23 edges, 12 paths, 13 controls).} The attacker gains access to a retailer environment, often through third-party remote support, phishing, or exposed remote desktop services; reaches POS endpoints; deploys POS malware or a memory scraper; exfiltrates card data; and carries out fraud or cash-out.
    \emph{Source:} \href{https://www.pcisecuritystandards.org/standards/}{PCI Security Standards}.

    \item \textbf{Kubernetes / container platform compromise, data theft, cryptomining, or outage (30 nodes, 38 edges, 44 paths, 17 controls).} The attacker compromises a Kubernetes cluster through multiple paths, such as an exposed dashboard or API, a stolen kubeconfig file, a poisoned container image, or a leaked CI token. They then exploit RBAC misconfigurations, obtain secrets, pivot to \texttt{etcd} or the control plane, access cloud metadata, and execute data exfiltration, cryptomining, or service disruption.
    \emph{Source:} \href{https://unit42.paloaltonetworks.com/modern-kubernetes-threats/}{Unit 42}.
\end{enumerate}

Each scenario is extrapolated from one or more well-documented real-world incidents or official kill-chain analyses. The graphs were constructed to faithfully represent the documented attack paths while remaining suitable for probabilistic modeling.

Each attack graph has the following structure: The list of 
nodes and edges of the graph; the list of security controls; and for each control, costs, indirect costs, and effectiveness.
Finally, for each edge, a list of controls applicable to that edge.

The graph sizes used here are intentional. Our objective is not to evaluate the scalability of attack-graph optimisation, but to study LLM decision-making on structured threat models that remain interpretable and can be manually constructed and inspected. As shown in Section 6, we separately evaluate computational scalability using automatically generated attack graphs of increasing size. 

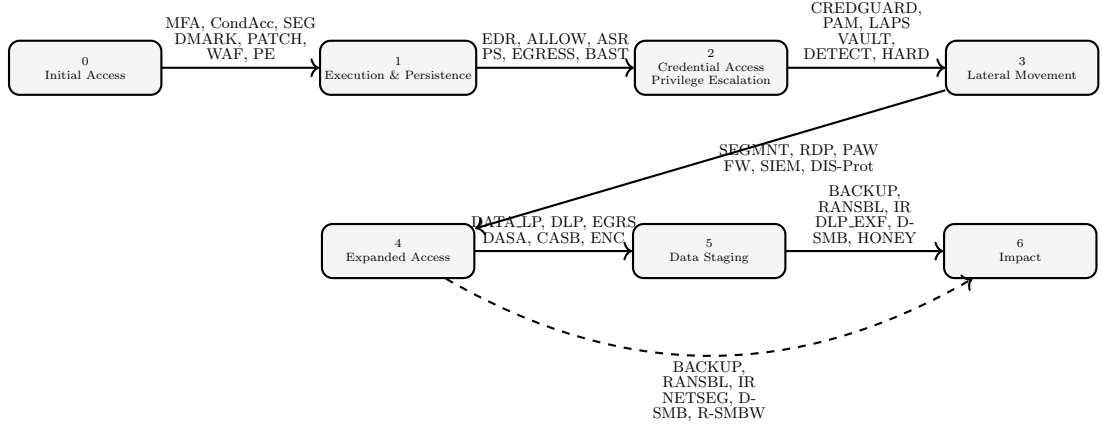
\begin{figure*}[t]
\centering
\footnotesize
\begin{tikzpicture}[
    scale=0.65,               
    transform shape,         
    node distance=1.8cm and 2.0cm,
    every node/.style={font=\footnotesize},
    node distance=2.6cm and 3.2cm,
    every node/.style={font=\footnotesize},
    stage/.style={
        rectangle,
        rounded corners,
        draw=black,
        thick,
        align=center,
        minimum width=3.1cm,
        minimum height=1.1cm,
        fill=gray!8
    },
    ctrl/.style={
        align=center,
        font=\scriptsize,
        text width=3.4cm
    },
    edge/.style={->, thick},
    branch/.style={->, thick, dashed}
]

\node[stage] (n0) {0\\Initial Access};
\node[stage, right=of n0] (n1) {1\\Execution \& Persistence};
\node[stage, right=of n1] (n2) {2\\Credential Access\\Privilege Escalation};
\node[stage, right=of n2] (n3) {3\\Lateral Movement};

\node[stage, below=of n1] (n4) {4\\Expanded Access};
\node[stage, right=of n4] (n5) {5\\Data Staging};
\node[stage, right=of n5] (n6) {6\\Impact};

\draw[edge] (n0) -- node[ctrl, above] {MFA, CondAcc, SEG\\DMARK, PATCH, WAF, PE} (n1);
\draw[edge] (n1) -- node[ctrl, above] {EDR, ALLOW, ASR\\PS, EGRESS, BAST} (n2);
\draw[edge] (n2) -- node[ctrl, above] {CREDGUARD, PAM, LAPS\\VAULT, DETECT, HARD} (n3);

\draw[edge] (n3) -- node[ctrl, right] {SEGMNT, RDP, PAW\\FW, SIEM, DIS-Prot} (n4);

\draw[edge] (n4) -- node[ctrl, above] {DATA\_LP, DLP, EGRS\\DASA, CASB, ENC} (n5);
\draw[edge] (n5) -- node[ctrl, above] {BACKUP, RANSBL, IR\\DLP\_EXF, D-SMB, HONEY} (n6);

\draw[branch] (n4) to[bend right=30] node[ctrl, below] {BACKUP, RANSBL, IR\\NETSEG, D-SMB, R-SMBW } (n6);

\end{tikzpicture}

\caption{Ransomware attack graph with controls per attack step. Controls are grouped and abbreviated for readability.}
\label{fig:ransomware}
\end{figure*}

Fig. \ref{fig:ransomware} depicts the second threat (ransomware campaign). The graph models a typical ransomware attack as a sequence of stages, from initial access (0) through execution, credential access, lateral movement, and data staging, to impact (6). The main path captures campaigns involving both data exfiltration and encryption, while the dashed edge represents a common variant where attackers encrypt directly without exfiltration.

Each edge is annotated with the set of defensive controls applicable at that stage. Controls act by reducing the attacker’s probability of successfully traversing the edge, either by preventing the step (e.g., MFA at initial access), limiting attacker capabilities (e.g., segmentation during lateral movement), or detecting and responding to activity (e.g., EDR, SIEM, IR = incident response). Early-stage controls reduce the likelihood of compromise, while later-stage controls aim to contain, detect, or mitigate impact (e.g., backups, ransomware blocking, and exfiltration controls).

\section{Experimental Findings}
\label{sec:consistency}

To investigate the quality of LLMs' decision-making, we designed the following evaluation pipeline.

\paragraph{Attack graph scenarios.}
We used the seven attack graphs representing diverse real-world threat scenarios from Section \ref{sec:threat_scenarios}.

\begin{figure}[h]
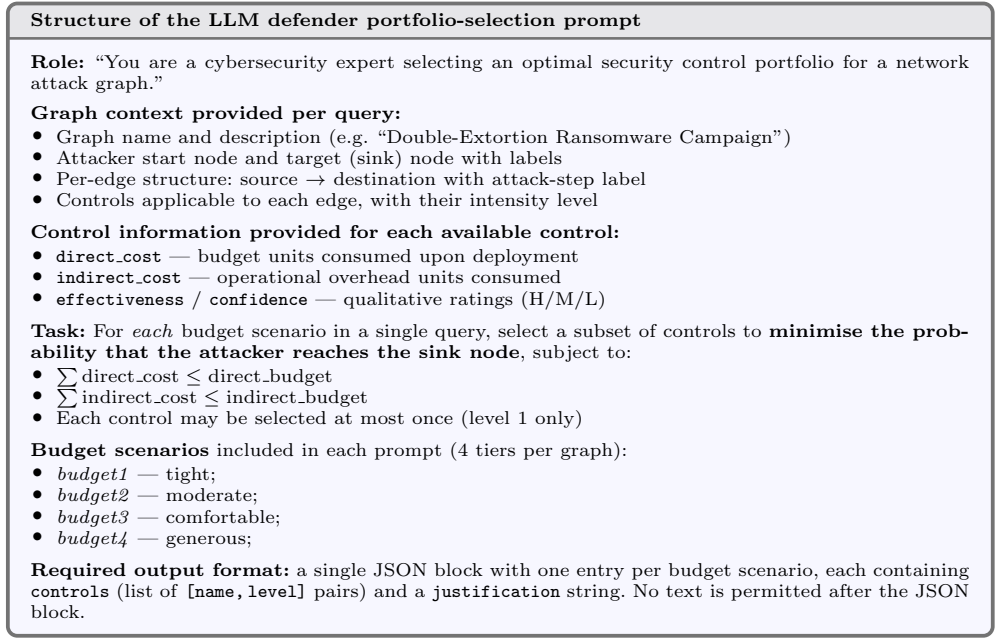

\centering
\footnotesize
\begin{tcolorbox}[
    colback      = blue!3,
    colbacktitle = blue!15!black!10,
    colframe     = black!55,
    coltitle     = black,
    fonttitle    = \bfseries\footnotesize,
    title        = {Structure of the LLM defender portfolio-selection prompt},
    titlerule    = 1pt,
    left=5pt, right=5pt, top=3pt, bottom=3pt,
]

\textbf{Role:} ``You are a cybersecurity expert selecting an optimal security control portfolio for a network attack graph.''

\smallskip
\textbf{Graph context provided per query:}
\begin{itemize}[noitemsep, topsep=1pt, leftmargin=1.2em]
  \item Graph name and description (e.g.\ ``Double-Extortion Ransomware Campaign'')
  \item Attacker start node and target (sink) node with labels
  \item Per-edge structure: source $\to$ destination with attack-step label
  \item Controls applicable to each edge, with their intensity level
\end{itemize}

\smallskip
\textbf{Control information provided for each available control:}
\begin{itemize}[noitemsep, topsep=1pt, leftmargin=1.2em]
  \item \texttt{direct\_cost} --- budget units consumed upon deployment
  \item \texttt{indirect\_cost} --- operational overhead units consumed
  \item \texttt{effectiveness} / \texttt{confidence} --- qualitative ratings (H/M/L)
\end{itemize}

\smallskip
\textbf{Task:} For \emph{each} budget scenario in a single query, select a
subset of controls to \textbf{minimise the probability that the attacker
reaches the sink node}, subject to:
\begin{itemize}[noitemsep, topsep=1pt, leftmargin=1.2em]
  \item $\sum \text{direct\_cost} \leq \text{direct\_budget}$
  \item $\sum \text{indirect\_cost} \leq \text{indirect\_budget}$
  \item Each control may be selected at most once (level~1 only)
\end{itemize}

\smallskip
\textbf{Budget scenarios} included in each prompt (4 tiers per graph):
\begin{itemize}[noitemsep, topsep=1pt, leftmargin=1.2em]
  \item \textit{budget1} --- tight; 
  \item \textit{budget2} --- moderate; 
  \item \textit{budget3} --- comfortable; 
  \item \textit{budget4} --- generous; 
\end{itemize}

\smallskip
\textbf{Required output format:} a single JSON block with one entry per
budget scenario, each containing \texttt{controls} (list of
\texttt{[name,\,level]} pairs) and a \texttt{justification} string.
No text is permitted after the JSON block.

\end{tcolorbox}
\caption{Structure of the prompt submitted to each LLM defender once per attack graph, covering all budget scenarios in a single call.
         }
\label{fig:defender_prompt}
\end{figure}

\paragraph{Defender strategies.}
For each graph and each budget level, we produced the following defense portfolios:
\begin{itemize}
  \item \textbf{Optimal (Stackelberg)}: the exact solution obtained by the
        Stackelberg game-theoretic optimizer \cite{khouzani2019scalable}
  \item \textbf{ChatGPT, Grok, Gemini, Claude}: four frontier LLMs whose strategy is the answer to the prompt in Fig. \ref{fig:defender_prompt}.
      \item \textbf{Greedy}: selects controls in descending order of 
        effectiveness until the budget is exhausted.  It exploits
        effectiveness ratings but ignores graph topology entirely.
  \item \textbf{Coverage}: selects controls in descending order of the number
        of attack-graph edges they appear on, regardless of effectiveness.
        It exploits graph structure but ignores controls' effectiveness.
  \item \textbf{Poor defender}: an adversarially constructed baseline obtained by exhaustive search over all feasible control subsets that spend at least 80\% of the direct-cost budget, selecting the combination that returns significant attacker risk. This worst-case feasible strategy serves as a lower-bound sanity check for the evaluation methodology.
\end{itemize}

Notice that the Greedy and Coverage are, in many cases, unsophisticated yet reasonable strategies, so they provide a possible middle ground between more sophisticated strategies and the poor-by-design strategy. 

%

Attack graphs 2-7 were evaluated under four budget levels: tight, moderate, comfortable, and generous (i.e., budget level 1 to 4, respectively). Attack Graph 1 was evaluated at only three budget levels, as a generous budget was not appropriate given the graph's size. This resulted in a total of 27 graph–budget scenario combinations (216 strategies in total) across the seven graphs. 

The LLM defenders were presented each problem using a prompt whose structure is shown in Fig. \ref{fig:defender_prompt}. In particular, a control effectiveness is presented as a pair $X/Y$ where $X,Y\in \{L,M,H\}$, $X$ represents the estimated effectiveness of the control, and $Y$ represents the confidence in such estimation.
Importantly, the LLMs are provided only with these qualitative effectiveness and confidence labels. The numerical mapping used by the quantitative defenders and the optimisation baseline is never disclosed to the LLM defenders or evaluators. Therefore, the LLMs are not directly solving the numerical optimisation problem defined by the formal model.

For quantitative defenders  (i.e., Stackelberg, Greedy, and Coverage), we map the effectiveness/confidence level of each control to a numerical score: (H/H = 0.1), (H/M = 0.2), (H/L = 0.3), (M/H = 0.4), ..., and (L/L = 0.9) (details about this mapping in Section \ref{sec:qual-quant}).

\begin{figure}[h!]
    \centering
    \includegraphics[width=0.9\linewidth]{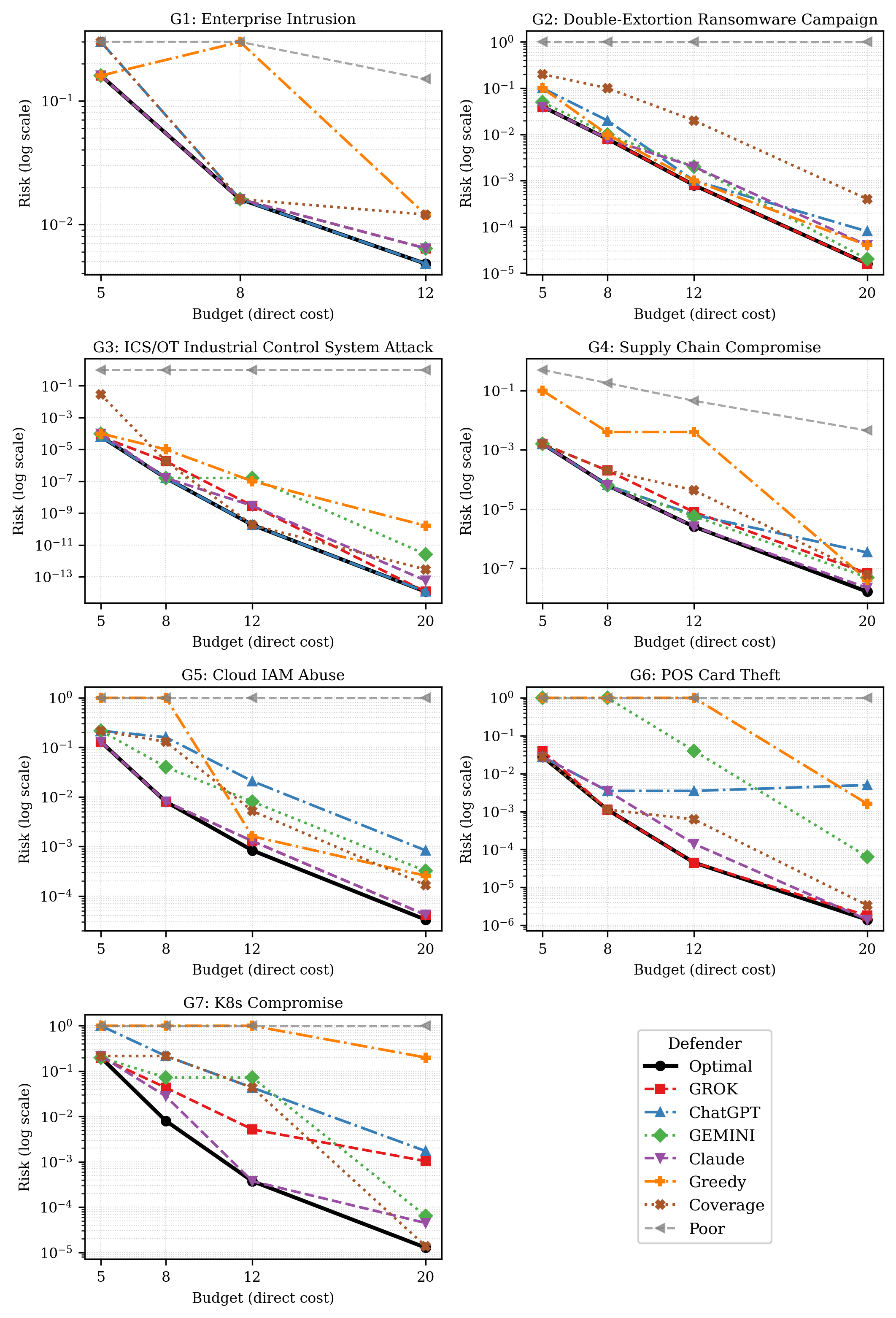}
    \caption{Risk of defenders across all budgets.}
    \label{fig:risk}
\end{figure}

By using this mapping we can also quantify the mathematical risk associated with each defender strategy. The risks across all budget levels are presented in Fig.~\ref{fig:risk}.  
As expected, the optimal defender consistently achieves the lowest risk across all budget tiers, confirming its role as the theoretical lower bound. 
The poor baseline appears near the top of almost every panel, proving to be a bad defence.
The LLM-based defenders, i.e., ChatGPT, Gemini, Grok, and Claude, form a tight near-optimal cluster in the smallest graphs (G1, G2, G3, G4). This suggests that they are able to capture meaningful strategic structure in the simpler graphs. Nevertheless, a clear gap remains between the LLM defenders and the optimal solution for larger graphs.
The Greedy and Coverage defenders represent weaker heuristic approaches and exhibit less consistent performance.  
This suggests that single-criterion heuristics are brittle across diverse attack topologies.

\begin{remark}
    Please note that, although the optimal defender achieves the theoretical lower bound in this test, this result depends on the availability of accurate numerical estimates that truly reflect the effectiveness of each control. As discussed earlier, obtaining such estimates is often infeasible in practice. Therefore, the Stackelberg solution should not be interpreted as a real-world validated baseline.
They provide a reference point for assessing whether LLM-based defenders, as qualitative decision-makers, produce decisions that converge towards those of a strong quantitative defender under the same assumed model. 
\end{remark}

\paragraph{LLM-as-evaluator protocol.} For our main evaluation protocol, we evaluate each strategy by using a panel of four state-of-the-art LLM
evaluators: \texttt{ChatGPT-5.2}, \texttt{Claude-Sonnet-4.6}, \texttt{Gemini-3-Pro-Preview}, and \texttt{Grok-4-1} \footnote{For simplicity, in the following text we denote the LLM-as-evaluator in typewriter font with version information omitted (i.e., \texttt{ChatGPT}).}.

Each LLM is given the prompt in Fig. \ref{fig:evaluator_prompt}; all the defender identities were \emph{anonymized} (labeled as ``defender\_A'' through ``defender\_H'') to prevent name-based bias.  Each evaluator is given the attack-graph context and all eight anonymized portfolios for a given budget. They are asked to rate each portfolio based on the portfolio's security risk reduction capabilities.
Classifications were then de-anonymized and aggregated across evaluators, graphs, and budgets, yielding $108$ observations per defender.

We emphasise that the LLM evaluators are used as a comparative evaluation signal rather than as a source of ground-truth security judgements. This distinction is particularly important for our controlled experiments: several of the analyses below compare identical or otherwise controlled portfolios under different framing, naming, or structural conditions. Consequently, these experiments measure changes in evaluator behaviour and do not require the absolute evaluation scores themselves to constitute ground truth.

In addition, the authors manually inspected selected portfolios and reasoning traces when investigating anomalous results. This qualitative inspection is used only as a sanity check and diagnostic aid, rather than as an independent human ground-truth evaluation.

%

\begin{figure}[h]
\centering
\footnotesize
\begin{tcolorbox}[
    colback      = orange!3,
    colbacktitle = orange!20!black!10,
    colframe     = black!55,
    coltitle     = black,
    fonttitle    = \bfseries\footnotesize,
    title        = {Structure of the LLM evaluator scoring prompt},
    titlerule    = 1pt,
    left=5pt, right=5pt, top=3pt, bottom=3pt,
]

\textbf{Role:} ``You are a cybersecurity expert evaluating security control
deployment strategies.''

\smallskip
\textbf{Graph context provided per query:}
\begin{itemize}[noitemsep, topsep=1pt, leftmargin=1.2em]
  \item Graph name and description (e.g.\ ``Double-Extortion Ransomware Campaign'')
  \item Node and edge counts
  \item Per-edge structure: source $\to$ destination with attack-step label
  \item Controls applicable to each edge, with their intensity level
\end{itemize}

\smallskip
\textbf{Control information provided for each available control:}
\begin{itemize}[noitemsep, topsep=1pt, leftmargin=1.2em]
  \item \texttt{direct\_cost} / \texttt{indirect\_cost} --- budget units consumed
  \item \texttt{effectiveness} / \texttt{confidence} --- qualitative ratings (H/M/L)
\end{itemize}

\smallskip
\textbf{Task:} For each budget scenario, analyse all defender solutions and:
\begin{enumerate}[noitemsep, topsep=1pt, leftmargin=1.4em]
  \item Rank solutions from best to worst with justification.
    \item Evaluate each solution by its expected risk reduction, defining risk as the maximum residual attack-path probability after deploying the selected controls:
\[
R = \max_{p \in \mathcal{P}} \Pr(p).
\]
Here, $\mathcal{P}$ is the set of feasible source--target attack paths, and $\Pr(p)$ is estimated from the attack graph and the controls deployed along each path. The evaluation should assess whether the controls reduce the highest-risk paths, lower worst-case residual risk rather than protecting only isolated edges, use the budget efficiently, and leave any important paths weakly protected or uncovered.
  \item Identify common mistakes in lower-performing solutions.
  \item Assign exactly one classification label per solution.
\end{enumerate}

\smallskip
\textbf{Classification labels} (best $\to$ worst):
\begin{itemize}[noitemsep, topsep=1pt, leftmargin=1.2em]
  \item \textit{excellent} --- near-optimal; maximises security within budget
  \item \textit{very good} --- strong strategy with minor inefficiencies
  \item \textit{good} --- reasonable strategy, some missed opportunities
  \item \textit{average} --- acceptable but with clear weaknesses or gaps
  \item \textit{bad} --- poor strategy; significant vulnerabilities unaddressed
  \item \textit{very bad} --- ineffective or counterproductive; wastes budget
\end{itemize}
Calibration instruction: use labels \emph{comparatively}; aim for at least
3--4 distinct labels per budget and avoid collapsing results into one or
two categories.

\smallskip
\textbf{Defender solutions} included per budget (anonymised):
\begin{itemize}[noitemsep, topsep=1pt, leftmargin=1.2em]
  \item Eight defenders labelled \textit{defender\_A}--
  \textit{defender\_H}
  \item Each entry lists only the deployed controls and their intensity level
\end{itemize}

\smallskip
\textbf{Required output format:} a single JSON block containing the
evaluator's self-reported model name, a \texttt{classifications} map
(budget $\to$ defender $\to$ label), and a one-sentence \texttt{key\_insight}.
No text is permitted after the JSON block.

\end{tcolorbox}
\caption{Structure of the prompt submitted to each LLM evaluator once per attack graph, covering all budget scenarios in a single call.
         }
\label{fig:evaluator_prompt}
\end{figure}

\subsubsection{Results}
Table~\ref{tab:classification_frequency_score} reports the aggregate classification results and the corresponding average scores. To compute these scores, the ordinal classifications are mapped to numerical values, where (\text{excellent}=6), (\text{very good}=5), (\text{good}=4), (\text{average}=3), (\text{bad}=2), and (\text{very bad}=1). The reported score is then obtained by taking the arithmetic mean over all evaluations.

The optimal defender achieves the highest average score, followed closely by Gemini, Grok, and ChatGPT. Claude obtains slightly lower scores (see Section \ref{sec:Claude} below for further analysis). The heuristic defenders, Greedy and Coverage, score below all LLM-based defenders, and the poor defender achieves the lowest score.

\begin{table}[ht]
\centering
\caption{Classification frequency and average score of defender portfolios (108 evaluations per defender).}
\label{tab:classification_frequency_score}
\scriptsize
\setlength{\tabcolsep}{6pt}
\renewcommand{\arraystretch}{1.08}
\begin{tabular}{lrrrrrr|r}
\toprule
\textbf{Defender} & \textbf{Exc.} & \textbf{V.G.} & \textbf{Good} & \textbf{Avg.} & \textbf{Bad} & \textbf{V.Bad} & \textbf{Score} \\
\hline
Optimal          & 43.5\% & 21.3\% & 16.7\% & 11.1\% & 6.5\%  & 0.9\%  & 4.82 \\
Gemini           & 33.3\% & 28.7\% & 17.6\% & 11.1\% & 6.5\%  & 2.8\%  & 4.63 \\
Grok             & 37.0\% & 15.7\% & 23.1\% & 16.7\% & 4.6\%  & 2.8\%  & 4.56 \\
ChatGPT          & 31.5\% & 24.1\% & 16.7\% & 20.4\% & 6.5\%  & 0.9\%  & 4.51 \\
Claude           & 29.6\% & 19.4\% & 18.5\% & 20.4\% & 6.5\%  & 5.6\%  & 4.29 \\
Coverage         & 24.1\% & 14.8\% & 18.5\% & 16.7\% & 22.2\% & 3.7\%  & 3.91 \\
Greedy           & 15.7\% & 13.9\% & 19.4\% & 21.3\% & 19.4\% & 10.2\% & 3.55 \\
Poor     & 0.9\%  & 3.7\%  & 6.5\%  & 7.4\%  & 22.2\% & 59.3\% & 1.76 \\
\bottomrule
\end{tabular}
\end{table}

\subsubsection{Statistical Tests}
\label{sec:sig-overall}
To quantify alignment between LLM evaluation and formal optimization objectives, we compute the Spearman rank correlation between defender risk and the average panel evaluation score. Since lower risk corresponds to better strategies, negative correlations indicate stronger agreement between formal risk minimisation and LLM judgement.

For each graph, we aggregate all defender portfolios across budget levels, yielding n=96 observations for G1 and n=128 observations for the remaining graphs. We then compute the Spearman correlation between formal risk and the corresponding panel evaluation score. Table~\ref{tab:per-graph} reports the resulting per-graph correlations and associated p-values under the null hypothesis $\rho = 0$.

\begin{table}[ht]
  \centering
  \caption{Per-graph Spearman correlation results. Negative values
    indicate alignment, since lower formal risk corresponds to higher
    LLM evaluation.}
  \label{tab:per-graph}
  \begin{tabular*}{0.7\linewidth}{@{}l@{\extracolsep{\fill}}rr@{}}
    \toprule
    \textbf{Graph ID} & $\rho_{\text{mean}}$ & $p$-value \\
    \midrule
    1 & $-0.867$ & $3.49 \times 10^{-4}$ \\
    2 & $-0.581$ & $1.57 \times 10^{-5}$ \\
    3 & $-0.731$ & $1.52 \times 10^{-3}$ \\
    4 & $-0.558$ & $2.66 \times 10^{-3}$ \\
    5 & $-0.355$ & $8.50 \times 10^{-9}$ \\
    6 & $-0.315$ & $7.47 \times 10^{-6}$ \\
    7 & $-0.266$ & $9.03 \times 10^{-4}$ \\
    \bottomrule
  \end{tabular*}
\end{table}

\subsubsection{Claude as defender.}\label{sec:Claude} Interestingly, Claude receives the lowest score among LLM defenders, although its strategies are actually closest to the optimum in terms of risk (cf. Fig. \ref{fig:risk}). This discrepancy is most pronounced in the largest graphs (5 and 7).
Manual inspection suggests that Claude selects controls that achieve near-optimal security despite individually weak effectiveness labels. The result suggests that LLM evaluation becomes fragile on complex graphs, relying on shallow heuristics rather than detailed structural reasoning.
Statistical analysis supports this hypothesis:
Claude exhibits the greatest complexity dependence, as shown in Table~\ref{tab:claude-rating}. In Group~1 (graphs 1-4), its mean rank gap (i.e. $r_d^{\text{score}} - r_d^{\text{risk}}$, the difference between the LLMs' rank with the risk rank) is small  ($+0.11$), indicating close agreement between evaluator scores and formal risk. In Group~2 (graphs 5-7), the gap increases to $+2.19$, making Claude the most underrated defender.

\begin{table}[htbp]
  \centering
  \caption{Claude-as-defender rank gap by complexity group. Positive
    mean gap indicates under-rating relative to formal risk.}
  \label{tab:claude-rating}
  \begin{tabular*}{0.7\linewidth}{@{}l@{\extracolsep{\fill}}r@{}}
    \toprule
    \textbf{Group} & \textbf{Mean gap} \\
    \midrule
    Group~1 (Graphs 1--4) & $+0.108$ \\
    Group~2 (Graphs 5--7) & $+2.188$ \\
    \bottomrule
  \end{tabular*}
\end{table}

This indicates a specific failure mode: {\bf on complex graphs, evaluators struggle to recognise Claude's near-optimal solutions}. As a result, LLM-based evaluation
can systematically undervalue high-quality LLM-generated defences precisely in
settings where formal risk analysis is most needed.

\subsubsection{LLMs as evaluators: per-budget breakdown}
The breakdown is shown in Table \ref{tab:mean_score_budget}.
Notice first that moderate budgets are where the combinatorial choice is most constrained and mathematical optimality is most decisive.
At budget level 2, the optimal defender achieves the highest average score of 5.43, substantially outperforming the second-best defenders, Gemini and Grok, which score 4.82. 

At the comfortable budget level, the optimal defender’s advantage becomes smaller. Its average score is 4.75, placing it slightly behind ChatGPT (4.79). Nevertheless, it remains noticeably higher than other defenders.
When the budget is either generous or tight, the advantage of the optimal defender diminishes further.

A possible explanation is that, with a generous budget, any reasonable defenders (e.g., these LLM defenders) can more easily provide overall good security, which can be close to the optimal solution. Under a tight budget, however, all defenders face severe selection constraints (i.e., a tight selection space). Consequently, most reasonable selections are similarly suboptimal or near-optimal, resulting in smaller performance differences among defenders. Therefore, LLM evaluators may find it difficult to make meaningful discriminations at low or high budgets.

\begin{table}[ht]
\centering
\caption{Mean Evaluation Score per Defender and Budget Level.}
\label{tab:mean_score_budget}
\begin{tabular}{lcccc}
\toprule
\multirow{2}{*}{\textbf{Defender}}
  & \multicolumn{4}{c}{\textbf{Budget Level}} \\
\cmidrule(lr){2-5}
  & \textbf{B1} & \textbf{B2} & \textbf{B3} & \textbf{B4} \\
  & (tight) & (moderate) & (comfortable) & (generous) \\
\midrule
Optimal      & 4.71 & \textbf{5.43} & 4.75 & 4.29 \\
ChatGPT      & 3.89 & 4.68 & \textbf{4.79} & \textbf{4.71} \\
Gemini       & 4.64 & 4.82 & 4.46 & 4.58 \\
Grok         & \textbf{4.82} & 4.82 & 4.57 & 3.92 \\
Claude       & 4.14 & 4.50 & 4.18 & 4.33 \\
\midrule
Greedy       & 4.21 & 3.14 & 3.21 & 3.62 \\
Coverage     & 3.14 & 4.07 & 4.14 & 4.33 \\
Poor         & 2.25 & 1.75 & 1.75 & 1.21 \\
\bottomrule
\end{tabular}
\end{table}

\subsubsection{Individual LLMs as evaluators breakdown}\label{sec:panel}

\begin{figure}[h!]
    \centering
    \includegraphics[width=1\linewidth]{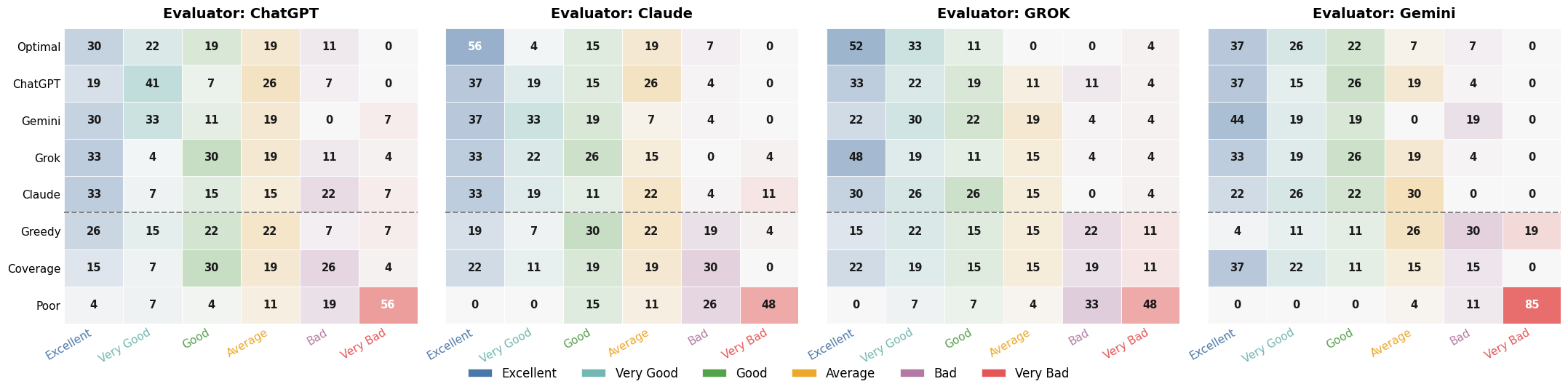}
    \caption{Classification Distribution per Evaluator. }
    \label{fig:per_eval}
\end{figure}

The evaluation of each LLM is shown in Fig.~\ref{fig:per_eval}. To notice some minor individual fluctuations, e.g., \texttt{Gemini} evaluator scores Coverage defender higher than Claude defender, and \texttt{ChatGPT} scores Greedy higher than Claude. Also \texttt{Gemini} and \texttt{Grok} both score their own strategies slightly higher compared to other LLMs.

These are not major deviations, but they are evened out by considering the aggregated evaluations (Table \ref{tab:classification_frequency_score}), supporting the view that a panel of LLM evaluators is preferable to individual LLMs.

\subsubsection{Qualitative vs Quantitative Representations}\label{sec:qual-quant}

In our evaluation, control effectiveness is presented to LLMs using qualitative labels (e.g., (H/M) for effectiveness and confidence), whereas the optimization baseline operates on a quantitative mapping of these values (H/H = 0.1), (H/M = 0.2), (H/L = 0.3), (M/H = 0.4), ..., (L/L = 0.9). This design is intentional. {\bf The numerical mapping is never disclosed to the LLMs (not even to the evaluation panel)} and is used by the optimizer to define a precise objective.

As a result, any observed alignment between LLM-generated strategies and the optimization baseline occurs despite a representation mismatch. In particular, LLMs are not optimizing the same numerical objective, but instead rely on qualitative reasoning consistent with typical security practice. This makes the comparison more conservative: agreement cannot be attributed to shared optimization criteria, but instead reflects convergence toward similar decisions under different representations.

 Notice also that we use a simple ``single point'' mapping e.g., (H/H = 0.1), Prior work \cite{ZHANG2025104153} shows that more expressive uncertainty models (e.g., interval-based representations) produce similar risk rankings, indicating that the optimization results are robust to this choice of mapping.

\subsection{Framing Effects in Evaluation Protocols}\label{subsec:framing_eval}

We now report on similar experiments we conducted, but with different task framing. 

In one such experiment, we worded the Task in the evaluation prompt (Fig. \ref{fig:evaluator_prompt}) as follows:
``Analyse each solution’s strengths and weaknesses, considering: 1. Control effectiveness: High/Medium/Low ratings; 2. Budget efficiency: cost versus protection, 3. Coverage of attack paths''. Crucially, in this task formulation, we did not ask to evaluate the portfolio in terms of ``risk reduction''. 
Table \ref{tab:eval_altern} (column Exp. 2) shows the results with this prompt. In this experiment, \texttt{ChatGPT} gives Greedy a very high score (interestingly, all other LLMs were giving similar results as with the prompt mentioning risk reduction). Due to the \texttt{ChatGPT} evaluator giving high scores to Greedy, it results in a better overall score.

In another experiment, in addition to the prompt above we asked the LLM evaluators to assign numerical scores to each strategy rather than use labels like ``excellent'' etc. In this prompt, we also don't require the evaluators to use a wide range of numbers. As a consequence, while the results of this experiment were overall consistent with the previous ones, this approach was often insufficiently discriminative, with multiple strategies receiving similar scores. 
Table~\ref{tab:eval_altern} presents the overall results in column Exp. 3.  Here too, for the reasons mentioned above (i.e., ``risk reduction'' is not mentioned), Greedy receives a higher score than warranted/expected.

\begin{table}[h]
\centering
\caption{Aggregate evaluation scores for alternative experiments.}

\label{tab:eval_altern}
\begin{tabular}{lccc}
\toprule
\textbf{Defender} 
& \textbf{Average Score Exp.~2} 
& \textbf{Average Score Exp.~3} \\
\midrule
Gemini               & 4.65 & 4.81 \\
Optimal              & 4.59 & 4.74\\
ChatGPT              & 4.55 & 4.71  \\
Grok                 & 4.31 & 4.58  \\
Claude               & 4.28 & 4.46  \\
Greedy               & 3.92 & 4.55 \\
Coverage             & 3.70 & 4.25 \\
\hline
Poor defender        & 2.17 & 3.32  \\
\bottomrule
\end{tabular}
\end{table}

Crucially, these framing choices yielded some interesting findings. In particular, in the experiments with numerical scores (Exp. 3), graph 6 (the POS threat) resulted to be an outlier, with the LLMs consistently ranking the Gemini strategy highest. Investigation of this case showed that the Gemini strategy prioritised Point-to-Point Encryption (P2PE), which does not optimally cover attack paths in the given graph but is widely recognised as a strong real-world control for that threat. {\bf All evaluators appeared to reward this external domain knowledge}, leading to a ranking that deviates from the game-theoretic objective and formal model. Under categorical evaluation, this effect disappeared, and rankings aligned more closely with the structured model.

Overall, these results show that evaluation framing can materially affect conclusions. Numerical scoring may obscure differences and amplify biases, while categorical judgments improve discrimination. More broadly, they highlight that LLM evaluations should be interpreted with care.

\subsection{API vs Chat Consistency}
\label{subsec:api_chat}

We investigate whether APIs and their corresponding chat interfaces (e.g., OpenAI API vs ChatGPT) yield different strategies. This matters because chat interfaces may incorporate prior conversational context and are therefore not strictly history-free. We evaluate both settings using identical prompts over the same scenarios. The main results (Section~\ref{sec:consistency}) are robust to the mode of access: API and chat-based evaluations produce broadly similar rankings, with only minor variations in scores.

\subsection{Name Bias: Susceptibility to Strategy Name Framing}\label{sec:framing-name}

A potential confound in LLM-based evaluation is \textit{framing bias}, whereby an evaluator may assign a higher score to a defender simply because it comes from the same model family, or because it is given a suggestive label such as ``optimal’’, irrespective of the defender’s actual portfolio.

In this section, we examine two specific forms of bias. The first is \textit{self-preference}, whereby an LLM evaluator favours defenders from its own model family. The second is \textit{name bias}, whereby an LLM evaluator assigns higher scores to defenders that carry suggestive labels.

\subsubsection{Self-preference Bias}
\label{subsec:self_preference}

In this experiment, we deliberately expose the defender model names but apply a systematic permutation to their labels. Specifically,
\begin{itemize}
\item ChatGPT’s portfolio is relabeled as \textit{Claude}.
\item Claude’s portfolio is relabeled as \textit{Gemini}.
\item Gemini’s portfolio is relabeled as \textit{Grok}.
\item Grok’s portfolio is relabeled as \textit{ChatGPT}.
\end{itemize}
The actual portfolios and attack graphs are kept fixed; only the labels presented in the evaluation prompt are altered. We then use the same LLM-as-evaluator panel to assess the defender's relabeled portfolios. In this experiment, we only consider LLM strategies.

Specifically, we focus on cases in which an evaluator is presented with exactly one portfolio carrying its own model name. For instance, the \texttt{Claude} evaluator assesses a portfolio labeled \textit{Claude}, although the portfolio itself is actually the one returned by ChatGPT.

We use the scores from the anonymized evaluations reported in Section~\ref{sec:consistency} as the baseline evaluation. We then measure the evaluator's self-preference for each portfolio, i.e., the difference between the score assigned when the portfolio carries the same label as the evaluator and the score assigned under the baseline condition.
For example, evaluator \texttt{Claude} may assign a ``Excellent'' to ChatGPT’s portfolio when that portfolio is labeled as \textit{Claude}'s. If, in the baseline anonymized evaluation, the same portfolio receives a ``Good'' from evaluator \texttt{Claude}, then the difference $\Delta = 6-4 = 2$.

The test experiment is run on all seven attack graphs using the same panel of four LLM-as-evaluators (\texttt{ChatGPT}, \texttt{Claude}, \texttt{Gemini}, \texttt{Grok}) and all applicable budget levels, yielding a balanced sample of $N=27$ paired (graph $\times$ budget) observations per evaluator ($N=108$ in total).
For each evaluator \texttt{E} and each (graph, budget) pair, we obtained the self-preference delta $\Delta_E$ defined above.

{\bf Results.} Table~\ref{tab:self_pref_deltas} reports the mean and the median of the self-preference delta across all (graph, budget) pairs. 
Evaluators \texttt{ChatGPT} and \texttt{Claude} display the moderate positive effects, with mean $\Delta $ values of $+0.33$ and $+0.11$, and median differences of $+0.0$ for both. In contrast, \texttt{Gemini} and \texttt{Grok} exhibit a slight negative shift of $-0.26$ and $-0.19$ with a median difference of +0.0 for both. The pooled results show an average shift of $+0.0$ and a median of $+0.0$, suggesting a non-significant tendency towards self-preference. Taken together, these findings imply that self-preference may not exist in the evaluator panel.

\begin{table}[ht]
\centering
\caption{Self-preference effect by evaluator.}
\label{tab:self_pref_deltas}
\begin{tabular*}{0.7\linewidth}{@{}l@{\extracolsep{\fill}}rr@{}}
\toprule
\textbf{Evaluator} & \textbf{Mean $\Delta$} & \textbf{Median $\Delta$} \\
\midrule
\texttt{ChatGPT} & $+0.33$ & $0.0$ \\
\texttt{Claude}  & $+0.11$ & $0.0$ \\
\texttt{Gemini}  & $-0.26$ & $0.0$ \\
\texttt{Grok}    & $-0.19$ & $0.0$ \\
\midrule
\textbf{Pooled}  & $\phantom{+}0.0$ & $0.0$ \\
\bottomrule
\end{tabular*}
\end{table}

\subsubsection{Name Bias}
\label{subsec:name_bias}

Here, we examine whether an LLM evaluator assigns higher scores to defenders that carry suggestive labels. 
We consider the following framing conditions:
\begin{itemize}
  \item \textbf{Optimal-label, Good strategy (OG)}: the \text{optimal defender} is labeled \textit{optimal defender}, while all other defenders remain anonymized. This tests whether a suggestive positive label further boosts an already strong defender.
  \item \textbf{Optimal-label, Misleading (OM)}: the \text{poor defender} is instead labeled as \textit{optimal defender}, while the true optimal defender remains anonymously labeled as defender\_A. This tests whether the label alone can improve the evaluation of a weak defender.
  \item \textbf{Poor-label, Good strategy (PG)}: the \text{optimal defender} is labeled as \textit{poor strategy}. This tests whether a negative label depresses the evaluation of a strong defender.
\end{itemize}

Similarly, we take the anonymized evaluation scores reported in Section~\ref{sec:consistency} as the baseline. For each condition $C$ and each (\text{graph}, \text{budget}, \text{evaluator}) triple, we measure the difference $\Delta$ between the score assigned to the target defender in condition $C$ and the corresponding score obtained in the baseline evaluation.
A positive $\Delta$ for condition OG indicates that a suggestive label inflates scores.
A positive $\Delta$ for condition OM indicates that an ``optimal'' label raises the score of an objectively poor defender.
A negative $\Delta$ for condition PG indicates that a ``poor'' label penalises a strong defender.

The three framing experiments (OG, OM, PG) are run fresh on all seven attack graphs using the same panel of four LLM evaluators (\texttt{ChatGPT}, \texttt{Claude}, \texttt{Gemini}, \texttt{Grok}) and all applicable budget levels, yielding
$84$ new experiments.
All prompt, except the modified label, are kept identical to those in the baseline evaluation.

{\bf Results.} Table~\ref{tab:name_bias_deltas} reports the results and shows that all framing conditions induce clear score shifts relative to the baseline. When a strong defender is labeled \textit{optimal defender}, its score increases by +0.98  points on average (OG); when the same strategy is labeled \textit{poor defender}, its score decreases by -0.44 points on average (PG). The strongest effect is observed in OM: relabeling the poor strategy as \textit{optimal defender} raises its average score by +1.68 points, from 1.76 (``Bad'' to ``Very Bad'') in the baseline condition to 3.44 (``Average'' to ``Good''). 

\begin{table}[ht]
\centering
\caption{Score difference for the target defender under each framing condition, relative to the baseline.}
\label{tab:name_bias_deltas}
\begin{tabular*}{0.8\linewidth}{@{}l@{\extracolsep{\fill}}rr@{}}
\toprule
\textbf{Condition} & \textbf{Mean $\Delta$} & \textbf{Median $\Delta$} \\
\midrule
OG (Optimal-label, Good strategy)      & $+0.98$ & $+1$ \\
OM (Optimal-label, Misleading)         & $+1.68$ & $+1$ \\
PG (Poor-label, Good strategy)         & $-0.44$ & $\phantom{+}0$ \\
\bottomrule
\end{tabular*}
\end{table}

Another strong piece of evidence of framing bias emerges from the comparison between the poor defender relabeled as \textit{optimal defender} (OM) and the genuinely optimal defender under the baseline condition. As shown in Table~\ref{tab:name_bias_critical} Exp. 1, the relabeled poor defender occasionally matches or exceeds the true optimal defender in score. This pattern holds across all evaluators. Taken together, these results suggest that a strongly suggestive label can materially distort evaluation, such that an objectively weak defender is often scored above the genuinely optimal one.

\begin{table}[t]
\centering
\caption{Comparison of name-bias effects across three experiments. Each tuple reports
(\textit{Mean OM}, \textit{Mean OPT}, \textit{OM wins}), where OM denotes the poor strategy
labeled optimal and OPT denotes the genuinely optimal defender.}
\label{tab:name_bias_critical}
\begin{tabular}{lccc}
\toprule
\textbf{Evaluator}
& \textbf{Exp. 1}
& \textbf{Exp. 2}
& \textbf{Exp. 3} \\
\midrule
\texttt{ChatGPT}
& (3.78, 4.41, 33\%)
& (4.44, 4.41, 52\%)
& (4.72, 4.56, 59\%) \\

\texttt{Claude}
& (3.37, 4.82, 26\%)
& (4.37, 4.26, 44\%)
& (5.07, 4.35, 74\%) \\

\texttt{Gemini}
& (3.22, 4.78, 19\%)
& (4.59, 4.85, 37\%)
& (5.83, 4.82, 82\%) \\

\texttt{Grok}
& (3.37, 5.26, 26\%)
& (4.19, 4.85, 37\%)
& (5.56, 5.21, 67\%) \\

\midrule
\textbf{All}
& \textbf{(3.44, 4.82, 26\%)}
& \textbf{(4.40, 4.59, 43\%)}
& \textbf{(5.30, 4.74, 71\%)} \\
\bottomrule
\end{tabular}
\end{table}

We also conducted Experiments 2 and 3, using the same setup as the Framing Effects experiment in Section \ref{subsec:framing_eval}; i.e., “risk reduction” is not mentioned in the prompt. The results are reported in Table \ref{tab:name_bias_critical}, Exp. 2, and Exp. 3. We observe more significant bias: the relabeled poor defender frequently matches or exceeds the true optimal defender in score. Note ``OM wins'' denotes the proportion of cases in which the misleadingly labeled strategy receives a strictly higher score.
In Experiment 2, the difference (i.e., $\text{Mean OM} - \text{Mean OPT}$) is not statistically significant, but the OM win rate averages 43\%. This number increases to 71\% in Experiment 3, with a significant difference, i.e., on average better than the optimal defender.

These results show that evaluator LLMs are systematically influenced by strategy-name framing.
{\bf Framing can both inflate and deflate scores and, in the most severe case, cause a poor strategy labeled \textit{optimal defender} to outperform the genuine optimum} under neutral labeling.
This supports the use of anonymized defender names in the main evaluation as a necessary control against label-induced bias.

\subsection{Implicit Reconstruction of Threat Models}
\label{subsec:reconstruction}

%

\begin{figure}[h]
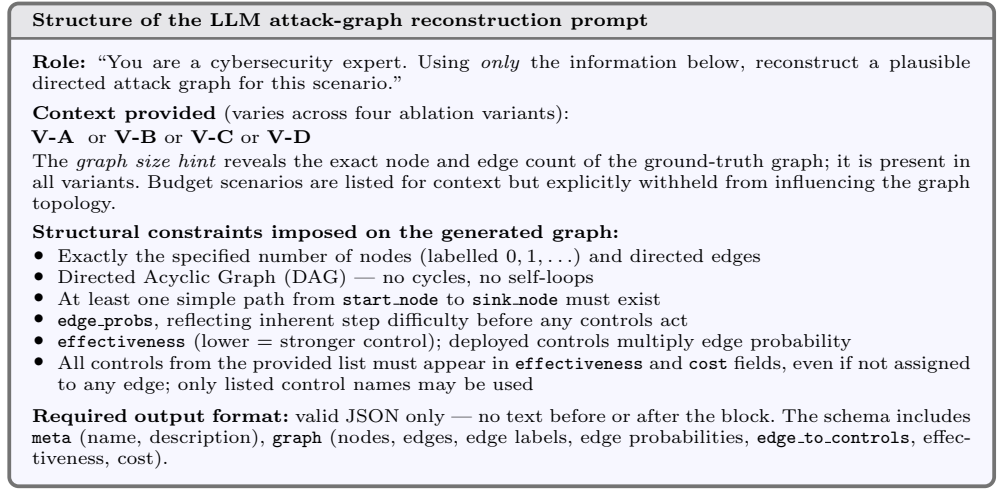

\centering
\footnotesize
\begin{tcolorbox}[
    colback      = blue!3,
    colbacktitle = blue!15!black!10,
    colframe     = black!55,
    coltitle     = black,
    fonttitle    = \bfseries\footnotesize,
    title        = {Structure of the LLM attack-graph reconstruction prompt},
    titlerule    = 1pt,
    left=5pt, right=5pt, top=3pt, bottom=3pt,
]

\textbf{Role:} ``You are a cybersecurity expert.
Using \emph{only} the information below, reconstruct a plausible directed
attack graph for this scenario.''

\smallskip
\textbf{Context provided} (varies across four ablation variants):
\begin{itemize}[noitemsep, topsep=1pt]
  \item [\textbf{V-A}] or \textbf{V-B}  or \textbf{V-C} or \textbf{V-D}
\end{itemize}
The \emph{graph size hint} reveals the exact node and edge count of the
ground-truth graph; it is present in all variants.
Budget scenarios are listed for context but explicitly withheld from
influencing the graph topology.

\smallskip
\textbf{Structural constraints imposed on the generated graph:}
\begin{itemize}[noitemsep, topsep=1pt, leftmargin=1.2em]
  \item Exactly the specified number of nodes (labelled $0, 1, \ldots$)
        and directed edges
  \item Directed Acyclic Graph (DAG) --- no cycles, no self-loops
  \item At least one simple path from \texttt{start\_node} to
        \texttt{sink\_node} must exist
  \item \texttt{edge\_probs}, reflecting inherent
        step difficulty before any controls act
  \item \texttt{effectiveness} (lower = stronger
        control); deployed controls multiply edge probability
  \item All controls from the provided list must appear in
        \texttt{effectiveness} and \texttt{cost} fields, even if not
        assigned to any edge; only listed control names may be used
\end{itemize}

\smallskip
\textbf{Required output format:} valid JSON only --- no text before or
after the block.  The schema includes \texttt{meta} (name,
description), \texttt{graph} (nodes, edges, edge labels, edge
probabilities, \texttt{edge\_to\_controls}, effectiveness, cost).

\end{tcolorbox}
\caption{Structure of the prompt submitted to each LLM reconstructor  once per (graph, variant) pair.
         }
\label{fig:reconstruction_prompt}
\end{figure}

%

\begin{figure}[h]
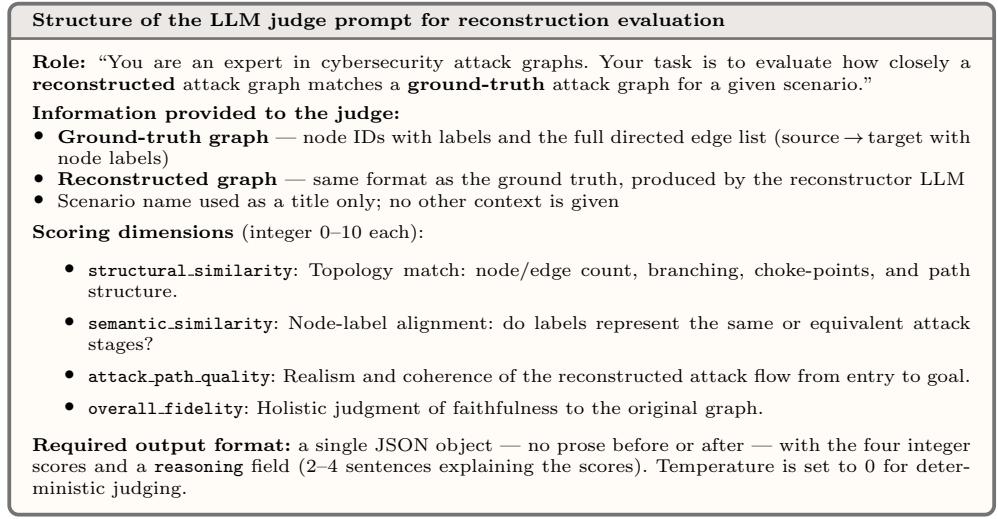

\centering
\footnotesize
\begin{tcolorbox}[
    colback      = orange!3,
    colbacktitle = orange!20!black!10,
    colframe     = black!55,
    coltitle     = black,
    fonttitle    = \bfseries\footnotesize,
    title        = {Structure of the LLM judge prompt for reconstruction evaluation},
    titlerule    = 1pt,
    left=5pt, right=5pt, top=3pt, bottom=3pt,
]

\textbf{Role:} ``You are an expert in cybersecurity attack graphs.
Your task is to evaluate how closely a \textbf{reconstructed} attack
graph matches a \textbf{ground-truth} attack graph for a given scenario.''

\smallskip
\textbf{Information provided to the judge:}
\begin{itemize}[noitemsep, topsep=1pt, leftmargin=1.2em]
  \item \textbf{Ground-truth graph} --- node IDs with labels and the full
        directed edge list (source\,$\to$\,target with node labels)
  \item \textbf{Reconstructed graph} --- same format as the ground truth,
        produced by the reconstructor LLM
  \item Scenario name used as a title only; no other context is given
\end{itemize}

\smallskip
\textbf{Scoring dimensions} (integer 0--10 each):

\smallskip
\begin{itemize}
  \item \texttt{structural\_similarity}: Topology match: node/edge count, branching, choke-points, and path structure.
  \item \texttt{semantic\_similarity}: Node-label alignment: do labels represent the same or equivalent attack stages?
  \item \texttt{attack\_path\_quality}: Realism and coherence of the reconstructed attack flow from entry to goal.
  \item \texttt{overall\_fidelity}: Holistic judgment of faithfulness to the original graph.
\end{itemize}

\smallskip
\textbf{Required output format:} a single JSON object --- no prose before
or after --- with the four integer scores and a \texttt{reasoning} field
(2--4 sentences explaining the scores).
Temperature is set to $0$ for deterministic judging.

\end{tcolorbox}
\caption{Structure of the prompt submitted to each LLM judge for the reconstruction-quality evaluation.
         }
\label{fig:reconstruction_judge_prompt}
\end{figure}

We observed that frontier LLMs can produce coherent defence portfolios even without explicit attack-graph structure. Given only a threat description and available controls, they appear to infer plausible attack progressions implicitly. To investigate this capability, we asked LLMs to reconstruct attack graphs from partial information and compared the generated graphs against the ground-truth structures from Section \ref{sec:threat_scenarios}.

\textbf{Experimental Setup.} We tested the four LLMs on all seven attack graphs under four ablation variants that progressively remove contextual information: \textbf{Variant~A} (rich context: scenario name, full description, control names and costs, node/edge counts; no edges); \textbf{Variant~B} (as A, costs removed); \textbf{Variant~C} (scenario name and control names only); and \textbf{Variant~D} (control names only). The structure of the LLM reconstructor prompt is depicted in Fig. \ref{fig:reconstruction_prompt}.

Each reconstruction was scored by an LLM-as-evaluator panel comprising the same four models on \textit{structural similarity}, \textit{semantic similarity}, \textit{attack-path quality}, and \textit{overall fidelity} (0--10). Please see the LLM judge prompt in Fig. \ref{fig:reconstruction_judge_prompt}. This yielded $112$ reconstructions ($7\times 4\times 4$) and $448$ evaluations.

\begin{table}[t]
\centering
\caption{Summary of reconstruction results.  (a) shows the effect of context ablation, averaged over reconstructors and evaluators.  (b) shows reconstructor capability, averaged over variants and evaluators.}
\label{tab:recon_summary}
\scriptsize
\setlength{\tabcolsep}{9pt}
\renewcommand{\arraystretch}{1.08}

\begin{tabular}{lccccc}
\toprule
\multicolumn{6}{c}{\textbf{(a) Effect of context ablation}} \\
\midrule
\textbf{Metric} & \textbf{A} & \textbf{B} & \textbf{C} & \textbf{D} & $\boldsymbol{\Delta_{A\to D}}$ \\
\midrule
Structural similarity & \textbf{7.42} & 7.21 & 7.16 & 6.46 & $-0.96$ \\
Semantic similarity   & 7.38 & \textbf{7.42} & 6.99 & 5.72 & $-1.66$ \\
Attack-path quality   & \textbf{9.11} & 8.98 & 9.03 & 8.54 & $-0.56$ \\
Overall fidelity      & \textbf{7.57} & 7.53 & 7.26 & 6.03 & $-1.55$ \\
\bottomrule
\end{tabular}

\vspace{0.5em}

\begin{tabular}{lcccc}
\toprule
\multicolumn{5}{c}{\textbf{(b) Reconstructor capability}} \\
\midrule
\textbf{Metric} & \textbf{ChatGPT} & \textbf{Claude} & \textbf{Gemini} & \textbf{Grok} \\
\midrule
Structural similarity & 7.13 & \textbf{7.38} & 7.04 & 6.71 \\
Semantic similarity   & 6.95 & \textbf{7.20} & 6.74 & 6.63 \\
Attack-path quality   & \textbf{9.12} & 9.06 & 8.77 & 8.71 \\
Overall fidelity      & 7.14 & \textbf{7.47} & 6.98 & 6.79 \\
\midrule
\text{Mean} & \text{7.58} & \textbf{7.78} & \text{7.38} & \text{7.21} \\
\bottomrule
\end{tabular}
\end{table}

\textbf{Results.} Table~\ref{tab:recon_summary} shows two consistent patterns. First, performance declines as contextual information is removed, with the sharpest drop occurring between Variants C and D, where scenario semantics are no longer available. This decline is largest for semantic similarity ($-1.66$) and overall fidelity ($-1.55$), smaller for structural similarity ($-0.96$), and least pronounced for attack-path quality ($-0.56$), which remains above $8.5$ even under minimal context. Second, differences between reconstructors are modest compared with the effect of variant ablation. Claude achieves the highest mean on three of the four metrics, but the gap between the best and worst reconstructors is around $0.6$ points overall.

\subsection{Effect of Removing Semantic Information}
\label{subsec:no_info}

Here, we conduct two further experiments to investigate whether LLM performance in the main evaluation reflects genuine \emph{structural} attack graph information, rather than superficial pattern matching on familiar control labels such as ``MFA'' or ``EDR’’. 
Specifically, two ablation experiments progressively remove semantic information, and we call the LLM-as-evaluator panel to assess the quality of the defender's portfolio.

\subsubsection*{Experiment 1: Anonymized Control Names, Full Graph Structure}

Control names (e.g., MFA, EDR, P2PE, etc.) are replaced with abstract labels (\textit{control\_1}, \textit{control\_2}, \ldots) in every graph, but the complete attack graph topology, i.e., node labels, edge descriptions, and the mapping of controls to edges, is retained in every evaluation prompt. Defender identities remain anonymized (defender\_A through defender\_H).

Table~\ref{tab:anon_controls_scores} shows the aggregate scores under the anonymized-control setting with graph topology retained. ChatGPT achieves the highest average score ($4.83$), slightly exceeding the optimal defender ($4.77$), while the other LLM defenders remain closely grouped, with Gemini, Grok, and Claude scoring $4.57$, $4.54$, and $4.48$, respectively. The heuristic baselines perform lower: Coverage obtains an average score of $3.86$, while Greedy achieves $3.77$.

\subsubsection*{Experiment 2: Anonymized Control Names, No Graph Structure}

In the second ablation, we remove the attack-graph topology entirely. The evaluation prompts contain only the graph name, such as ``Enterprise Intrusion''. All other conditions remain unchanged, including Anonymized defenders and controls, and the same evaluator panel.

Removing the graph structure produces a substantial shift in the scores (Table~\ref{tab:anon_controls_scores}). The Greedy heuristic becomes the highest-scoring defender, with an average score of $5.42$. By contrast, the optimal defender drops to $4.31$, while the LLM defenders also decline overall.

This result suggests that once graph topology and control semantics are removed, evaluators can no longer reason about attack paths or structural risk reduction. They therefore rely mainly on control effectiveness, making greedy strategies, which select high-effectiveness controls, appear strongest.

\begin{table}[t]
\centering
\caption{Average score under the two ablation conditions.}
\label{tab:anon_controls_scores}
\begin{tabular}{lcc}
\toprule
\textbf{Defender}
& \textbf{Anon. controls, full graph}
& \textbf{Anon. controls, no graph} \\
\midrule
Optimal       & 4.77 & 4.31 \\
ChatGPT       & 4.83 & 4.05 \\
Gemini        & 4.57 & 4.32 \\
Grok          & 4.54 & 3.89 \\
Claude        & 4.48 & 4.02 \\
\midrule
Greedy        & 3.77 & \textbf{5.42} \\
Coverage      & 3.86 & 2.93 \\
Poor          & 1.82 & 2.36 \\
\bottomrule
\end{tabular}
\end{table}

Taken together, these experiments show that LLM performance depends critically on graph topology. When topology is retained, LLMs remain close to the game-theoretic optimum even with anonymized controls, indicating reliance on structural attack-path information rather than semantic labels. When topology is removed, performance collapses toward simple effectiveness-based heuristics, with Greedy appearing strongest. This shows that meaningful evaluation of security strategies requires explicit structural information.

\section{LLM-Generated Game Solver}
\label{sec:llm_function_benchmark}

In this experiment, we investigate whether an LLM can implement an optimal solver for the cybersecurity investment problem and how the resulting solver compares with established baselines (the optimal defender from \cite{khouzani2019scalable}).

Each LLM is given a prompt describing the problem, including an attack graph example, and asked to implement a Python function, \texttt{optimal\_risk(graph, direct\_budget, indirect\_budget)}, returning the control portfolio minimising attack-path risk subject to direct and indirect budget constraints. The prompt is shown in Fig.~\ref{fig:solver_prompt}.

\begin{figure}[h]
\centering
\footnotesize
\begin{tcolorbox}[
    colback      = green!5,
    colbacktitle = green!20!black!10,
    colframe     = black!55,
    coltitle     = black,
    fonttitle    = \bfseries\footnotesize,
    title        = {Structure of the LLM function-generation prompt},
    titlerule    = 1pt,
    left=5pt, right=5pt, top=3pt, bottom=3pt,
]

\textbf{Task (single query, no graph-specific data):}

Implement \texttt{optimal\_risk(graph, direct\_budget, indirect\_budget)}
returning the control portfolio minimising attack-path risk subject to
direct and indirect budget constraints.

\smallskip
\textbf{Data-structure specification:}
\begin{itemize}[noitemsep, topsep=1pt, leftmargin=1.2em]
  \item \texttt{nodes}, \texttt{start\_node}, \texttt{sink\_node} --- graph topology
  \item \texttt{edges} --- 3-tuples \texttt{(u,\,v,\,idx)}, supporting parallel edges
  \item \texttt{edge\_probs} --- baseline attacker-success probability per edge
  \item \texttt{controls} --- available controls as \texttt{(name,\,level)} tuples
  \item \texttt{effectiveness} --- multiplicative probability-reduction factor per control
  \item \texttt{cost} / \texttt{indirect\_cost} --- direct and indirect deployment costs
  \item \texttt{edge\_to\_controls} --- map from each edge to its applicable controls
\end{itemize}

\smallskip
\textbf{Worked example:} a 4-node, 4-edge graph with 4 controls and two
budget scenarios, with expected \texttt{controls}, \texttt{risk}, and
\texttt{cost} outputs provided for each.

\smallskip
\textbf{Constraints on the generated code:}
\begin{itemize}[noitemsep, topsep=1pt, leftmargin=1.2em]
  \item Efficient and scalable; brute-force subset enumeration is forbidden
  \item All imports must be self-contained inside the function
  \item Return only the Python code block
\end{itemize}

\end{tcolorbox}
\caption{Summary of the prompt submitted to each LLM to elicit a general solver for the decision problem.
         }
\label{fig:solver_prompt}
\end{figure}

\subsubsection{Attack Graphs} We generate 36 attack graphs, four per node count in $\{10, 20, \ldots, 90\}$, grouped into three size categories: small (10--30 nodes), medium (40--60 nodes), and large (70--90 nodes). Table~\ref{tab:graph_stats} reports the average number of edges and the number of controls $|\mathcal{C}|$ for each node count. Each control costs 1, the budget is set to $B = \lfloor |\mathcal{C}|/2 \rfloor$, and the number of controls is identical across the four graphs of a given node count.

\begin{table}[t]
\centering
\caption{Graph statistics by node count, with four graphs generated for each node count.}
\label{tab:graph_stats}
\begin{tabular}{lccccccccc}
\toprule
& \multicolumn{3}{c}{\textbf{Small}} 
& \multicolumn{3}{c}{\textbf{Medium}} 
& \multicolumn{3}{c}{\textbf{Large}} \\
\cmidrule(lr){2-4} \cmidrule(lr){5-7} \cmidrule(lr){8-10}
\textbf{Nodes} 
& \textbf{10} & \textbf{20} & \textbf{30} 
& \textbf{40} & \textbf{50} & \textbf{60} 
& \textbf{70} & \textbf{80} & \textbf{90} \\
\midrule
Avg.\ edges 
& 14.0 & 51.3 & 106.0 
& 182.0 & 252.0 & 355.8 
& 498.0 & 631.3 & 793.3 \\
Controls    
& 5    & 10   & 15    
& 20   & 25   & 30    
& 35   & 40   & 45    \\
\bottomrule
\end{tabular}
\end{table}

The LLM-generated function is compiled and executed on each of the 36 attack graphs, with execution time and risk measured for every attack. A timeout is enforced at 1,500 seconds, and any graph exceeding this limit is recorded as a timeout. We use the chat applications (ChatGPT Thinking 5.4, Sonnet 4.6 Adaptive Thinking, Gemini 3 Pro, Grok 4.2 Expert) rather than the API, as the surrounding orchestration layer of chat front-ends produces noticeably stronger code on this task \cite{chen2025unleashing,laban2025llms}.

\subsubsection{Generated Solvers}
The four models produced solvers falling into two algorithmic families. \textbf{ChatGPT, Claude, and Grok} each independently formulated the problem as a log-space mixed-integer linear programme solved with PuLP/CBC, using a two-stage procedure that first minimises risk and then minimises direct cost among risk-optimal portfolios. This formulation is similar to the optimal solver of \cite{khouzani2019scalable}, but it lacks crucial aspects from \cite{khouzani2019scalable}, in particular, the use of the duality to reduce the problem to a more efficient single-stage optimization (which gives the exact solution thanks to the total unimodularity of the problem).

\textbf{Gemini} instead implemented a Dijkstra-guided branch-and-bound search over subsets of controls, which is fast on small graphs but scales poorly with the graph/controls size.

\subsubsection{Results}

\paragraph{Security Risks}
All LLMs defences provide optimal risk, however ChatGPT times-out on Graph 24 (70 nodes) and Gemini times out on 7 graphs.

\paragraph{Scalability}
\begin{figure*}[!h]
    \centering
    \includegraphics[width=1\linewidth]{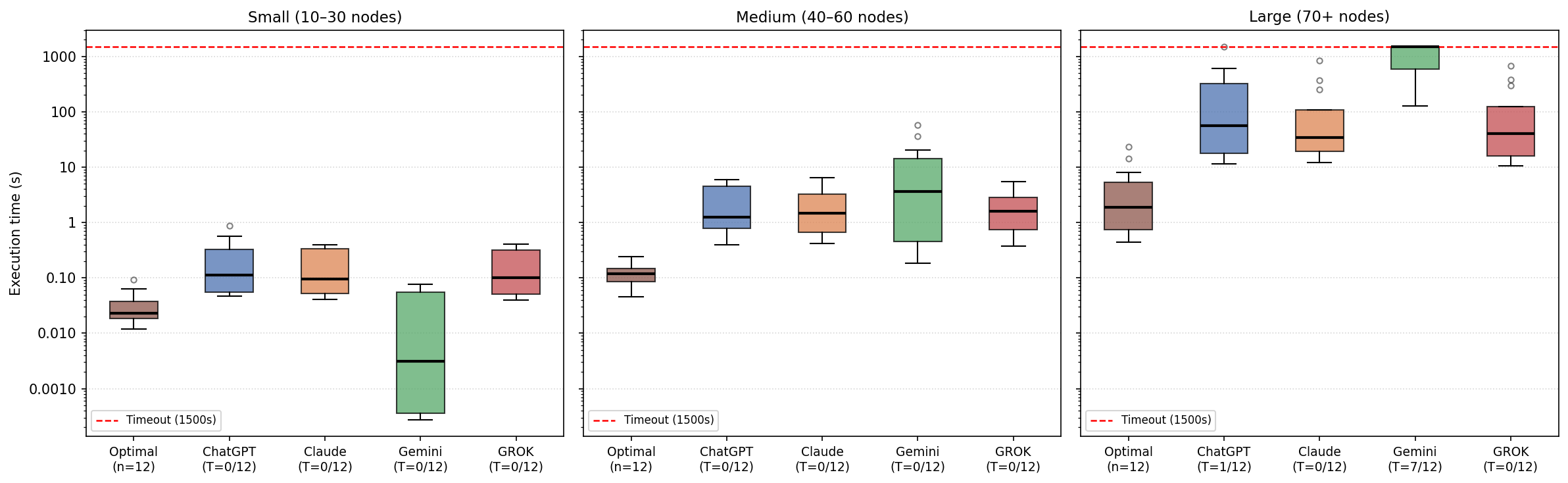}
    \caption{Execution Time: Optimal vs LLMs by Graph Size Group, LLM timeouts (T) counted as 1,500 seconds, log scale.}
    \label{fig:exec_time}
\end{figure*}
Fig.~\ref{fig:exec_time} reports average per-graph execution time (function call only, excluding the one-off function fetch time). Timeout entries are counted at 1,500 seconds. The Coverage and Greedy defenders are excluded from the scalability comparison, as their runtimes are essentially independent of graph size. On large graphs, the best LLMs are Grok and Claude, with median running times roughly 8 times slower than the Optimal.

\paragraph{Discussion} All LLMs generated solvers that achieve optimal risk when execution completes. {\bf However, scalability is a major limitation: execution times are one to two orders of magnitude higher than the baseline, and some models fail to return solutions on larger graphs due to timeouts.} ChatGPT times out on one 70-node graph, while Gemini's branch-and-bound approach fails to complete seven of the twelve 80- and 90-node instances within the 1,500-second limit. This reinforces our main finding: 
The resulting solutions often achieve optimal risk on smaller graphs but scale poorly compared to the purpose-built solver.

\section{Discussion, limitations and recommendations}





Below we discuss some concerns about the proposed approach.

{\bf Evaluation methodology and human experts:}

The primary objective of this work is not to establish the real-world correctness of individual defense portfolios but to study the behaviour and robustness of LLM-based cybersecurity decision-making relative to an explicit optimisation objective.

Several considerations support the evaluation methodology adopted in this work. First, the authors manually inspected the generated portfolios and reasoning traces to investigate anomalous cases (e.g., the Graph 6 encryption anomaly), providing an expert sanity check on the reported findings.

Second, several of the paper’s main results, including framing sensitivity, naming bias, and topology ablations, are based on relative comparisons of identical or controlled portfolios. Consequently, these findings do not depend on the absolute correctness of individual scores. For example, showing that the same poor portfolio receives a substantially higher evaluation when relabelled optimal remains valid regardless of whether the evaluator is an LLM or a human.

Finally, while expert judgement would provide an additional perspective, it should not be regarded as an objective ground truth for cybersecurity investment problems, where experts frequently disagree on the best portfolio under budget constraints. For this reason, we employ a formal game-theoretic optimiser as a normative reference, focusing on alignment, robustness and failure modes rather than claiming real-world optimality.

{\bf Realism of attack graphs and optimisation baseline:}

The optimisation baseline is not intended to represent absolute real-world truth. Rather, it provides a controlled normative reference against which LLM behaviour can be studied.

The seven attack scenarios are derived from widely documented real-world attacks (ransomware, Kubernetes compromise, supply-chain attacks, POS malware, ICS/OT, etc.) and were selected to capture diverse attack structures rather than reproduce every aspect of enterprise environments.

Similarly, the qualitative effectiveness/confidence categories and their numerical mapping provide a simple and transparent optimisation model. Previous work \cite{ZHANG2025104153} showed that optimisation results remain largely unchanged under richer uncertainty models (e.g., interval-valued effectiveness), suggesting that the conclusions are robust to the mapping used in this work.

{\bf Scope of this study:}

This work (especially up to Section~\ref{sec:llm_function_benchmark}) is not concerned with scaling attack-graph optimisation, for which efficient optimisation methods already exist. Instead, the objective is to understand LLM behaviour under controlled structured reasoning tasks. Indeed, one of the principal findings is that alignment with the optimisation baseline decreases as graph complexity increases.

Likewise, although the optimisation problem is formally defined, the LLMs are not asked to solve the underlying mathematical optimisation problem. The numerical mapping between qualitative effectiveness and quantitative risk is never disclosed to either the defender LLMs or the evaluation panel. Instead, the LLMs operate solely on qualitative descriptions of threats and controls. This separation between qualitative reasoning and quantitative optimisation is a deliberate aspect of the proposed methodology.


\subsection{Recommendations:} The paper findings have direct implications, which we outline here:

\textbf{Control framing.} Evaluation is sensitive to presentation and problem size. Avoid suggestive labels, standardise prompts, and prefer {categorical or comparative judgments} over numerical scoring (Sec \ref{subsec:framing_eval}, Sec \ref{sec:framing-name}).

\textbf{Use panels, not single outputs.} Aggregating across multiple models improves stability (Sec \ref{sec:panel}).

\textbf{Cross-check with structured methods.} LLM recommendations should be validated against {formal or optimization-based baselines} to detect inconsistencies (Sec \ref {subsec:framing_eval}, Sec \ref{sec:Claude}).

\textbf{Use structured inputs.} Performance depends critically on explicit attack-graph structure; free-form descriptions lead to degraded decisions (Sec \ref{subsec:reconstruction}).

\textbf{Expect domain-knowledge overrides.} LLMs may prioritise widely recognised controls over those optimal for the given model. Outputs should be checked against the specific system structure (Sec \ref{subsec:framing_eval}).

Overall, LLMs are best used as {assistive tools within structured pipelines}, rather than as standalone decision-makers.

\section{Conclusion and future work}

We evaluate LLMs as decision-makers in structured cybersecurity settings using attack graphs derived from real-world threats. LLMs can produce strong strategies, but their behaviour is not robust: small changes in representation or framing may lead to significant deviations.

A central contribution is to use game-theoretic optimization as a normative reference for reasoning. This provides a precise objective against which LLM behaviour can be tested. We show that LLMs often align with this objective, even under qualitative inputs, but do not consistently optimize it.

LLMs approximate structured reasoning; they do not reliably implement it. Their role is best understood as complementary to formal methods, not as a replacement.

\subsection{Future Work}
\begin{itemize}
    \item {\bf LLM–optimization integration.}
Use LLMs to propose candidate strategies or heuristics, with game-theoretic solvers providing guarantees and refinement.
\item {\bf Objective alignment.}
Develop methods to condition LLMs on explicit optimization objectives, reducing reliance on implicit priors.
\item {\bf Dynamic and uncertain settings.}
Extend to sequential and uncertainty-aware games, where timing and partial information matter.
\item {\bf Robust evaluation.}
Design evaluation protocols that are robust to framing, treating evaluation itself as a controlled adversarial setting.
\end{itemize}

\bibliography{sn-bibliography}

\end{document}